\documentclass[a4paper,12pt]{article}
\usepackage{amsmath}
\usepackage{ulem}
\usepackage{calc}
\usepackage{vmargin}
\usepackage{amstext}
\usepackage{graphicx}
\usepackage{amsfonts}
\usepackage{amssymb}
\usepackage{comment}

\begin{document}
	
	\title{\bf An Operator Approach to the Integration of Linear Differential Equations}
	
	\author{ \bf O.V. Kaptsov
		\\ Federal Research Center for Information and Computational Technologies,
		\\ Novosibirsk, Russia
		\\ Email: kaptsov@mail.ru}
	
	\date{}
	\maketitle
	
	\numberwithin{equation}{section}
	
	\begin{abstract}
		We develop an operator approach to the integration of linear differential equations based on intertwining relations between differential operators. Conditions for the existence of intertwining operators are obtained, and it is shown that, in low-order cases, the problem reduces to Riccati-type equations. The method is applied to linear partial differential equations, which makes it possible to construct their solutions. The linear Klein--Gordon equation is presented as an illustrative example.
	\end{abstract}
	

	\section{Introduction}
	
	Linear differential equations with variable coefficients arise in many problems of mathematical physics, including wave theory, quantum mechanics, spectral theory, and vibration theory. Of particular interest are constructive methods that establish connections between different differential operators, thereby allowing one to derive new equations and their solutions from already known ones. Such methods include group analysis \cite{Ovs}, factorization of differential operators \cite{Ince, InfeldHull}, Darboux transformations \cite{Matveev}, as well as various operator and algebraic approaches that are widely used, in particular, in supersymmetric quantum mechanics and in the theory of exactly solvable models \cite{Junker, CooperKhareSukhatme}.
	
	In the present paper, we employ an operator approach in which a linear differential equation is identified with the action of a differential operator.
	%from the ring $\mathcal{F}[\partial]$, where $F$ is a commutative ring equipped with a derivation $\partial$. 
	The central object of the approach is the intertwining relation
	\[
	MT = TL,
	\]
	which connects two operators $L$ and $M$ by means of a differential operator $T$. This relation defines a map from solutions of the equation $Lu=0$ to solutions of the equation $Mv=0$, and thus induces a natural equivalence relation on the set of linear differential operators. We show that the problem of constructing first-order intertwining operators leads to Riccati-type equations, which can be reduced to linear differential equations by a standard substitution.
	
	The operator approach also proves to be effective for linear partial differential equations. The use of commuting operators makes it possible to transfer results obtained for ordinary differential operators to the Klein--Gordon equation
	\[
	u_{tt} = u_{xx} + V(x)u,
	\]
	and to construct equations with new potentials together with their solutions. Such constructions arise naturally in problems of mathematical physics related to wave processes \cite{CooperKhareSukhatme, Br, Lan}. It should be noted that Euler was the first to propose such an approach and to apply it to problems in acoustics \cite{Euler}. At present, this approach is commonly referred to as the Darboux method. Mention should also be made of the largely forgotten article by Moutard on the transformation and integration of linear ordinary differential equations \cite{Moutard}.
	
\section{Intertwining Operators}

Let $\mathcal{F}$ be a commutative ring equipped with a derivation $\partial : \mathcal{F} \to \mathcal{F}$. We denote by $\mathcal{F}[\partial]$ the ring of differential operators of the form
\[
L = \sum_{i=0}^{n} a_i \partial^i, \qquad a_i \in \mathcal{F}.
\]
If $a_n \neq 0$, then the integer $n$ is called the order of the operator $L$.
Multiplication in the ring of operators is defined by the rule
\[
\partial f = f' + f \partial, \qquad f' = \partial(f).
\]
The ring $\mathcal{F}[\partial]$ is neither commutative nor a unique factorization domain.

We recall the notion of an intertwining operator. An operator $T \in \mathcal{F}[\partial]$ is said to intertwine an operator $L$ with an operator $M$ if
\begin{equation} \label{T}
	MT = TL.
\end{equation}
The operator $T$ is called an intertwining operator, and equality \eqref{T} is called the intertwining condition.

From now on, let $\mathcal{F}$ be the ring of germs of infinitely differentiable functions at a point $x_0 \in \mathbb{R}$.

\medskip
\noindent \textbf{Lemma 1.}
Let $L$ be an operator of the form
\[
L = a_n \partial^n + \dots + a_1 \partial + a_0.
\]
Then there exist an operator $T = \partial + s$, where $s \in \mathcal{F}$, and an operator $M$ of order $n$ such that the intertwining condition \eqref{T} is satisfied.

\medskip
\noindent {\it Proof.}
Let us represent the operator $M$ in the form
\[
M = b_n \partial^n + b_{n-1} \partial^{n-1} + \dots + b_0, \qquad b_i \in \mathcal{F},
\]
where $b_n(x_0) \neq 0$.
We expand the left-hand and right-hand sides of the intertwining condition.

The left-hand side of \eqref{T} can be written as
\[
(b_n \partial^n + \dots + b_0)(\partial + s) = b_n \partial^{n+1} + c_n \partial^n + \dots + c_0,
\]
where
\[
c_m = b_{m-1} + \sum_{k=m}^{n} b_k \binom{k}{m} s^{(k-m)}, \qquad m>0.
\]
The coefficient $c_0$ is given by
\[
c_0 = b_n s^{(n)} + b_{n-1} s^{(n-1)} + \dots + b_0 s.
\]

The right-hand side of \eqref{T} has the form
\[
a_n \partial^{n+1} + (a_n' + s a_n)\partial^n + \dots + (a_1' + s a_1)\partial + a_0' + s a_0.
\]
By comparing the coefficients, we obtain
\[
b_n = a_n,
\]
and the remaining coefficients $b_{n-1}, \dots, b_0$ are determined uniquely.

For $s$, we obtain an ordinary differential equation
\[
a_n s^{(n)} + b_{n-1} s^{(n-1)} + \dots + b_0 s = a_0' + s a_0.
\]
Given $n$ initial conditions, the solution of this equation depends smoothly on these data.

Let us consider two examples. Suppose that $L$ and $M$ are second-order operators:
\[
L = a_2 \partial^2 + a_1 \partial + a_0, \qquad 
M = b_2 \partial^2 + b_1 \partial + b_0.
\]
From the intertwining condition \eqref{T}, as described in Lemma~1, we find the coefficients
\[
b_2 = a_2, \qquad b_1 = a_2' + a_1, \qquad 
b_0 = a_0 + a_1' - s a_2' - 2 a_2 s'.
\]
The differential equation for $s$ takes the form
\[
a_2 s'' - 2 s s' a_2' - a_2' s^2 + s' a_2' + s' a_1 + s a_1' - a_0' = 0.
\]
Integrating this equation, we obtain a Riccati equation
\[
a_2 (s' - s^2) + a_1 s - a_0 = \lambda, \qquad \lambda \in \mathbb{R}.
\]
The substitution $s = -(\ln h)'$ reduces it to the linear equation
\begin{equation} \label{Eq a2}
	a_2 h'' + a_1 h' + (a_0 + \lambda) h = 0.
\end{equation}

Now consider a pair of third-order operators:
\[
L = a_3 \partial^3 + a_2 \partial^2 + a_1 \partial + a_0, \qquad
M = b_3 \partial^3 + b_2 \partial^2 + b_1 \partial + b_0.
\]
The intertwining condition yields the coefficients
\[
b_3 = a_3, \qquad b_2 = a_2 + a_3',
\]
\[
b_1 = a_1 + a_2' - s a_3' - 3 a_3 s',
\]
\[
b_0 = a_0 + a_1' + 3 s' s a_3 + a_3' s^2 - 2 s' a_3' - 2 s a_2' - 3 a_3 s'' - a_2' s,
\]
and leads to a third-order differential equation for $s$.
This equation admits a first integral and can be reduced to a second-order equation
\[
a_3 (s'' + s^3) - 3 a_3 s' s'' - a_2 s^2 + s' a_2 + s a_1 - a_0 = \lambda, \qquad \lambda \in \mathbb{R}.
\]
The substitution $s = -(\ln h)'$ reduces it to the linear equation
\[
a_3 h''' + a_2 h'' + a_1 h' + (a_0 + \lambda) h = 0.
\]
Thus, the equation for $s$ can be regarded as a second-order Riccati-type equation.

\medskip
\noindent \textbf{Lemma 2.}
Let there be given a linear differential equation of order $n$
\begin{equation} \label{Ln}
	L(u) = a_n u^{(n)} + \dots + a_1 u' + a_0 u = 0, \qquad a_i \in \mathcal{F},
\end{equation}
and let $h \in \mathcal{F}$ satisfy the equation
\begin{equation} \label{h}
	L h = \lambda h, \qquad \lambda \in \mathbb{R},
\end{equation}
with $h(x_0) \neq 0$.
Then the transformation
\[
w = u' - \frac{h'}{h} u
\]
maps solutions of equation \eqref{Ln} to solutions of some linear differential equation of order $n$ for the function $w$.

\medskip
\noindent {\it Proof.}
By assumption, $h \in \mathcal{F}$ satisfies equation \eqref{h}.
Let $u = h v$, where $v \in \mathcal{F}$. Then $v$ satisfies the equation
\[
a_n h v^{(n)} + \left(n a_n h' + a_{n-1} h\right) v^{(n-1)} + \dots + \left(a_n h^{(n)} + \dots + a_1 h' + a_0 h\right) v = 0.
\]
Using \eqref{h}, this equation can be rewritten in the form
\[
a_n v^{(n)} + \dots + \lambda v = 0.
\]
Differentiating this equation and setting $q = v'$, we obtain an $n$th-order equation
\[
a_n q^{(n)} + \tilde{a}_{n-1} q^{(n-1)} + \dots + \tilde{a}_0 q = 0, \qquad \tilde{a}_i \in \mathcal{F}.
\]
Evidently, this equation is satisfied by the function
\[
q = \left( \frac{u}{h} \right)'.
\]
Thus, the function $w = q h = u' - (\ln h)' u$ also satisfies a linear differential equation of order $n$.

\medskip
Let us briefly consider the factorization of a second-order operator
\[
L = F \partial^{2} + G \partial + H, \qquad F, G, H \in \mathcal{F}.
\]
We seek to represent $L$ in the form
\begin{equation} \label{L1L2}
	L = L_2 L_1,
\end{equation}
where $L_1 = \partial + s$, $L_2 = F \partial + p$, and $s, p \in \mathcal{F}$.
By comparing the coefficients, we obtain two relations:
\[
p = G - s F, \qquad H = F s' + s p.
\]
Thus, $L$ factorizes if
\begin{equation} \label{L2}
	L_2 = F \partial + (G - s F)
\end{equation}
and $s$ satisfies the Riccati equation
\[
F (s' - s^2) + s G - H = 0.
\]
The substitution $s = -(\ln h)'$ reduces this equation to the linear equation
\begin{equation} \label{hla}
	F h'' + G h' + H h = 0.
\end{equation}
We observe that equation \eqref{hla} is obtained from \eqref{Eq a2} for $\lambda = 0$. Factorization is used to construct operator intertwinings and transformations of differential equations \cite{CooperKhareSukhatme}.

\medskip
As shown above, the operator $T = \partial + s$ intertwines the operators
\[
L = F \partial^2 + G \partial + H, \qquad 
M = F \partial^2 + G_1 \partial + H_1,
\]
if $G_1 = G + F'$, $H_1 = H + G' - s F' - 2 s' F$, and $s$ satisfies the equation
\begin{equation} \label{Fs}
	F (s' - s^2) + G s - H = \lambda, \qquad \lambda \in \mathbb{R}.
\end{equation}
A direct calculation shows that the operator
\[
T^c = F \partial + (G - s F)
\]
intertwines $M$ with $L$. Formally, $T^c$ coincides with \eqref{L2}. However, in the present case $s$ satisfies \eqref{Fs}, whereas in the factorization case $s$ satisfies this equation only for $\lambda = 0$.

\medskip
We say that two operators $L$ and $M$ are equivalent if they are intertwined. It is easy to see that intertwining defines an equivalence relation on the set of operators. This leads to a broader notion of equivalence for differential equations than the one usually adopted \cite{Olver}.

\section{Linear Partial Differential Equations}

Let us now consider the ring $F_2$ of germs of smooth functions of two variables at a point $(x_0, y_0) \in \mathbb{R}^2$. Let $F_x$ and $F_y$ denote the rings of germs of smooth functions at the points $x_0$ and $y_0$, respectively. Two derivations, $\partial_x$ and $\partial_y$, act on $F_2$. The corresponding ring of differential operators is denoted by $F_2[\partial_x, \partial_y]$. The intertwining of operators $L, M \in F_2[\partial_x, \partial_y]$ is defined by formula~\eqref{T}, where $T \in F_2[\partial_x, \partial_y]$.

If $L_1 \in F_x[\partial_x]$ and $L_2 \in F_y[\partial_y]$, then these operators commute. Suppose that $L_1$ is intertwined with $M \in F_x[\partial_x]$ by means of an operator $T \in F_x[\partial_x]$, that is,
\[
MT = T L_1.
\]
Since $L_2 T = T L_2$, it follows that
\[
(M + L_2) T = T (L_1 + L_2).
\]
Consequently, if $u \in F_2$ satisfies the equation
\[
(L_1 + L_2) u = 0,
\]
then $v = T u$ is a solution of the equation $(M + L_2) v = 0$.

Consider the Klein--Gordon equation
\begin{equation} \label{utt}
	u_{tt} = u_{xx} + V(x) u, \qquad V(x) \in F_x.
\end{equation}
It corresponds to the operator $\partial_t^2 - \partial_x^2 - V(x)$.
As shown above, the operator $L_1 = \partial_x^2 + V(x)$ is intertwined with the operator $M = \partial_x^2 + V(x) + 2 (\ln h)''$ by means of $T = \partial_x - (\ln h)'$, provided that $h$ satisfies the equation
\begin{equation} \label{h2}
	h'' + (V(x) + \lambda) h = 0.
\end{equation}
Thus, the operator $\partial_t^2 - \partial_x^2 - V(x)$ is intertwined with $\partial_t^2 - \partial_x^2 - (V(x) + 2 (\ln h)'')$. This allows us to construct solutions of the equation
\begin{equation} \label{vtt}
	v_{tt} = v_{xx} + (V(x) + 2 (\ln h)'') v,
\end{equation}
provided that solutions of equations \eqref{utt} and \eqref{h2} are known.

Let us present several examples. Consider the wave equation
\[
u_{tt} = u_{xx},
\]
which has the general solution $u_0 = X(t+x) + Y(x-t)$, where $X$ and $Y$ are arbitrary functions.
Since $V(x) = 0$ in this case, equation \eqref{h2} takes the form
\begin{equation} \label{h0}
	h'' + \lambda h = 0.
\end{equation}
For $\lambda = 0$, equation \eqref{h0} has the solution
\[
h = c_0 + c_1 x, \qquad c_0, c_1 \in \mathbb{R}.
\]
In this case, equation \eqref{vtt} becomes
\begin{equation} \label{v2}
	v_{tt} = v_{xx} - \frac{2}{(x+b)^2} v, \qquad b = \frac{c_0}{c_1}.
\end{equation}
Accordingly, a solution is given by
\[
v = \partial_x (u_0) - \frac{u_0}{x+b}.
\]

If $\lambda = -k^2$, then a solution of equation \eqref{h0} is
\[
h = c_0 \sinh(kx) + c_1 \cosh(kx), \qquad c_0, c_1 \in \mathbb{R}.
\]
The corresponding equation \eqref{vtt} takes the form
\[
v_{tt} = v_{xx} + \frac{2k^2 (c_1^2 - c_0^2)}{[c_0 \sinh(kx) + c_1 \cosh(kx)]^2} \, v,
\]
and a solution is given by
\[
v = \partial_x (u_0) - k \frac{c_0 \cosh(kx) + c_1 \sinh(kx)}{c_0 \sinh(kx) + c_1 \cosh(kx)} \, u_0.
\]
For $\lambda = k^2$, the corresponding equation for $v$ and its solution are obtained from the above formulas by replacing hyperbolic functions with trigonometric ones.

Let us now consider equation \eqref{v2} together with the equation for $h$:
\begin{equation} \label{H}
	h'' + \left( \lambda - \frac{2}{(x+b)^2} \right) h = 0.
\end{equation}
If $\lambda = 0$, then a solution of equation \eqref{H} is
\[
h = c_1 (x+b)^2 + \frac{c_2}{x+b}, \qquad c_1, c_2 \in \mathbb{R}.
\]
For $\lambda < 0$, we obtain the solution
\[
h = c_1 e^{m x} \left( m^2 - \frac{m}{x+b} \right) + c_2 e^{-m x} \left( m^2 + \frac{m}{x+b} \right),
\]
where $m^2 = -\lambda$. This allows us to construct solutions of the equation
\[
w_{tt} = w_{xx} - 2 \left( \frac{1}{(x+b)^2} - (\ln h)'' \right) w.
\]

The equation
\[
v_{tt} = v_{xx} - \frac{2}{\sinh^2 x} v
\]
and the corresponding equation for $h$,
\[
h'' - \left( \frac{2}{\sinh^2 x} + A^2 \right) h = 0, \qquad A \in \mathbb{R},
\]
can be treated in a similar manner. For $A = 2$, a solution of the latter equation is
\[
c_1 (\cosh(2x) - 1) + c_2 \frac{\sinh(2x) (\cosh(2x) - 2)}{\cosh(2x) - 1}, \qquad c_1, c_2 \in \mathbb{R}.
\]
For an arbitrary $A > 0$, the explicit solution formula becomes rather cumbersome.

\medskip
\noindent \textit{Remark.}
This procedure can be continued as long as equation \eqref{h2} can be solved. Using Crum's formula, one can construct solutions of equations for $h$ from known solutions of equation \eqref{h0} corresponding to different values $\lambda_1, \dots, \lambda_n$.

\medskip
Separation of variables combined with the operator approach can also be used to construct solutions of equation \eqref{utt}. As an example, consider the equation
\begin{equation} \label{W}
	u_{tt} = u_{xx} + \left( -\frac{x^2}{4} + k \right) u, \qquad k \in \mathbb{R}.
\end{equation}
We seek a solution in the form $u = T(t) X(x)$. The functions $T$ and $X$ satisfy the equations
\[
T'' + \lambda T = 0, \qquad 
X'' + \left( k - \lambda - \frac{x^2}{4} \right) X = 0.
\]

The equation for $X$ is usually referred to as the Weber equation, and its solutions are well known \cite{WhittakerWatson}. In particular, if $k - \lambda = n + \frac{1}{2}$, where $n$ is a nonnegative integer, one of its solutions has the form
\[
X = (-1)^n e^{\frac{x^2}{4}} \frac{d^n}{dx^n} \left( e^{-\frac{x^2}{2}} \right).
\]
Subsequently, one can apply the operator approach to equation \eqref{W} using this known solution of the Weber equation or directly employ Crum’s determinant formula.

The equation
\[
u_{tt} = u_{xx} - \bigl( k + n(n+1) \wp(x) \bigr) u
\]
can be investigated in a similar manner, where $\wp$ is the Weierstrass elliptic function, $n$ is a positive integer, and $k \in \mathbb{R}$. After separation of variables, we arrive at the Lamé equation, whose solutions can be found in the classical literature \cite{WhittakerWatson}.

\section{Conclusion}

In this paper, we developed an operator approach to linear differential equations based on the intertwining relation $MT = TL$. We showed that the construction of a first-order intertwining operator of the form $T = \partial + s$ is essentially reduced to the solution of a Riccati-type equation for the coefficient $s$. This observation provides a unifying and structurally transparent framework for Darboux-type transformations.

From an algebraic perspective, the obtained results clarify the structure of factorizations and intertwinings in the ring $\mathcal{F}[\partial]$. From a constructive viewpoint, they provide an effective tool for generating new exactly solvable operators and potentials arising in mathematical physics. Possible directions for further research include the extension of the present approach to higher-order intertwining operators and the investigation of its applications to integrable systems and nonlinear partial differential equations. This approach carries over to the case of many variables.

\section*{Acknowledgments}
This work was carried out within the framework of the state assignment of the Ministry of Science and Higher Education of the Russian Federation for the Federal Research Center for Information and Computational Technologies.

The work was supported by the Krasnoyarsk Mathematical Center, funded by the Ministry of Science and Higher Education of the Russian Federation within the framework of the establishment and development of Regional Mathematical Research Centers (Agreement No.~075-02-2026-735).

\end{document}